\documentclass[aps,prl,twocolumn,showpacs,groupedaddress]{revtex4}

\usepackage{amsmath}
\usepackage{graphicx}
\usepackage{dcolumn}
\usepackage{bm}

\begin{document}

\title{High-order Time Expansion Path Integral Ground State}

\author{R. Rota, J. Casulleras, F. Mazzanti, and J. Boronat}
\affiliation{
 Departament de F\'\i sica i Enginyeria Nuclear, Campus Nord
B4-B5, Universitat Polit\`ecnica de Catalunya, E-08034 Barcelona, Spain}

\date{\today}

\begin{abstract}
The feasibility of path integral Monte Carlo ground state calculations with
very few beads using a high-order short-time Green's function expansion is
discussed. An explicit expression of the evolution operator which provides
dramatic enhancements in the quality of ground-state wave-functions is
examined. The efficiency of the method makes possible to remove the trial
wave function and thus obtain completely model-independent results still
with a very small number of beads. If a single iteration of the method is
used to improve a given model wave function, the result is invariably a
shadow-type wave function, whose precise content is provided by the
high-order algorithm employed. 
\end{abstract}

\maketitle


Quantum Monte Carlo path integral calculations provide a powerful
approach to many-body physics, both at zero~\cite{sarsa,galli,cuervo} 
and finite temperature~\cite{ceperley}.
They rely on using the classical action of the system in imaginary
time to define a path distribution probability, following Feynman's
approach. For systems with Bose statistics, the probability is positive
and the only approximation involved in the calculations comes from
the discretization of time, which prevents an exact evaluation of
the action. In path integral calculations, chains with large numbers
of time slices or beads are just a realization of these
ideas~\cite{chandler}. If an
accurate evaluation of the action for paths with large time steps
were possible, the complexity of path integral calculations, and in
particular ergodicity issues, would be greatly reduced. 

For ground-state calculations, the path integral approach (PIGS)~\cite{sarsa,galli,cuervo}  
provides
a method to systematically improve a trial wave function, by repeated
application of the evolution operator in imaginary time, which eventually
drives the system into the ground state~\cite{ciftja}. Here again, if an accurate
implementation of the evolution operator for large time steps were
available, it could be used to build variational wave functions of
very high quality.

In this letter, we focus on high-order short-time expansions of the
Green's function, which have been the subject of a series of works in
recent years~\cite{takahashi,suzuki,chin1}. This progress has led to a deep understanding of their properties
and provided various extremely accurate decompositions of the evolution
operator $\hat{U}= \exp (\varepsilon\: \hat{H})$, with $\hat{H}= \hat{T}+
\hat{V}$, expressed as products
of unitary operators of the basic components $\hat{U}= \exp (\varepsilon\:
\hat{T})$
and $\hat{U}= \exp(\varepsilon\: \hat{V})$. An important result is that the use
of a finite time step $\varepsilon$ can be renormalized in the potential.
This means that for a range of time steps $\varepsilon$, one is
in practice virtually free from finite time step errors. 

We make use of the decomposition schemes proposed
by Chin ~\cite{chin1,chin2} in order to evaluate the feasibility of PIGS
calculations with high-order propagators using very few beads ($N_b$). To this end,
we revisit the ground state of bulk superfluid $^4$He. We also  
explore the possibility of performing
ground-state calculations of boson systems without any model wave function, i.e., to take
as a starting point $\Psi=1$ and rely in the propagator's quality
to build the actual wave function exclusively from the Hamiltonian.
It is also interesting to note that if one does make
use of a trial wave function, the application of Chin's evolution
operator~\cite{chin1,chin2} produces a much enhanced model that actually
incarnates a shadow wave function~\cite{ciftja}. In this way, Chin's analysis
provides a deep understanding of the success of shadow-like wave
functions~\cite{vitiello,reatto}
and sheds light onto the actual mechanisms leading to this kind
of wave-functions.


Decompositions of the evolution operator preserving unitarity in the
form
\begin{equation}
\exp(\varepsilon\:(\hat{T}+\hat{V}))=\prod_{i=1}^{N}\exp(t_{i}\varepsilon
\hat{T}) \exp(v_{i}
\varepsilon \hat{V})+ {\cal O}(\varepsilon^{n+1}) \ ,
\label{expansion1}
\end{equation}
 are the starting point which have culminated in the algorithms due
to Chin that we test here. By means of a proper selection of the factorization
coefficients $\{ t_{i}\}$ and $\{ v_{i}\}$, any given order can be
achieved (the resulting expression is $n$th order since the effective
Hamiltonian is then $\hat{T}+\hat{V}+ {\cal O}(\varepsilon^{n})$). 
However, if some
of the $t_{i}$ factorization coefficients are negative, involving
diffusion processes backwards in time, the algorithm cannot be used
in the context of quantum Monte Carlo calculations. Fourth order algorithms,
for which Chin has worked out a complete characterization~\cite{chin1}, happen to
be somehow unique, as for only forward decompositions, the highest
order that can be achieved is four.  These developments result from  
a careful use of the expansion
\begin{eqnarray}
\lefteqn{\prod_{i=1}^{N} \exp(t_{i}\varepsilon \hat{T}) \exp(v_{i}\varepsilon
\hat{V})  = 
\exp(\varepsilon e_{T} \hat{T} + \varepsilon e_{V} \hat{V} +
\varepsilon^{2} e_{TV} } 
\label{expansion2} \\
& & \times [\hat{T},\hat{V}] 
 + \varepsilon^{3}e_{TTV}[\hat{T},[\hat{T},\hat{V}]]+
\varepsilon^{3}e_{VTV}[\hat{V},[\hat{T},\hat{V}]] )+ \ldots
\nonumber
\end{eqnarray}
 which allows to keep only the simplest term $[\hat{V},[\hat{T},\hat{V}]]$, 
 which produces a sort of renormalized potential
\begin{equation}
[\hat{V},[\hat{T},\hat{V}]] = \frac{\hbar^2}{m} \sum_{i=1}^{N} |\bm{F}_i|^2
\,
\label{conmutador}
\end{equation}
with $\bm{F}_i=\sum_{j\neq i}^{N} \bm{\nabla}_i V(r_{ij})$.
A remarkable characteristic of the resulting expansions is that they lead
to a continuous family of 4th order algorithms characterized by one free
parameter. It is interesting to note that with a proper choice of that
parameter ($t_1$ in Eq. \ref{algc}) one is able to fine tune the leading error term
of the propagator to any desired value, including changing sign in a
controlled way. Thus, one is enabled to minimize the value of the 4th order
error coefficient, and in doing so, even try to cancel, to the largest
possible extent, the contributions of the next orders. In practice, this
means that a particular value of the free parameter entering in the
algorithm produces exceedingly accurate and stable results. A different
strategy to improve the order of the propagator is the use of a
multiproduct expansion with a controlled violation of
positivity~\cite{zillich}.

Amongst the various decompositions proposed in Ref. \cite{chin1}, the particular
scheme chosen in this work is
\begin{equation}
e^{-\tau \hat{H}}=e^{-v_{0}\tau\: \hat{V}}e^{-t_{1}\tau\: \hat{T}}
e^{-v_{1}\tau\: \hat{W}} e^{-t_{2}\tau\: \hat{T}} e^{-v_{1}\tau\: \hat{W}} e^{-t_{1}\tau\: \hat{T}}
e^{-v_{0}\tau\: \hat{V}}
\label{algc}
\end{equation}
 with $\hat{W}= \hat{V} + (u_0/v_1) \tau^2 \, [\hat{V},[\hat{T},\hat{V}]]$. 
The various factorization coefficients
are all dependent on the single free parameter $t_{1}$~\cite{chin1}.
The range of possible values for $t_{1}$is $0\leq t_{1}\leq 0.5$
and experience shows that the optimal value is nearly independent
of $\tau=\varepsilon / N_b$. A similar decomposition has been recently used in path integral
Monte Carlo (PIMC) simulations at finite temperature 
showing high accuracy~\cite{sakkos}.



The operator splitting (\ref{algc}) allows for an estimation of the Green's
function $G(R,S,\tau)=\left\langle R|e^{-\tau H}|S\right\rangle$, which in
turn provides the  mechanism for building accurate
wave functions starting from a 
trial wave function, $\Psi(R)=\int G(R,S,\tau)\Psi_{\text m}(S)\: dS$. More
explicitly,
\begin{eqnarray}
\lefteqn{ \Psi(R) =  \int e^{-v_{0}\tau V(R)} e^{-\frac{(R-S_{1})^{2}}{4Dt_{1}\tau}}
e^{-v_{1}\tau W(S_{1})} 
e^{-\frac{(S_{1}-S_{2})^{2}}{4Dt_{2}\tau}}} \label{eq:abs} \\ \nonumber
& & \times  \, e^{-v_{1}\tau W(S_{2})}
e^{-\frac{(S_{2}-S_{3})^{2}}{4Dt_{1}\tau}}e^{-v_{0}\tau V(S_{3})}
\:\Psi_{\text m}(S_{3})\: dS_{1}dS_{2}dS_{3} \ .
\end{eqnarray}

Actually, Eq. \ref{eq:abs} sets the grounds for building wave functions
consistent with the decomposition of the propagator (\ref{algc}).
It has an intuitive content too: it states that the actual
value of $\Psi(R)$ should be taken as a weighted average of neighbor
shadow values $\Psi(S)$, with a weight given by a precise combination
of exponentials of the (finite-time-step renormalized) potential.
Moreover, it is also very suggesting to consider the double possibility that
Eq. \ref{eq:abs} offers, either as a powerful enhancement of an a priori
model wave function $\Psi_{\text m}(R)$ or directly as a tool to generate
the wave function directly from the Hamiltonian.

Equation \ref{eq:abs} can also be written in terms of relative rather than absolute
auxiliary coordinates, 
\begin{eqnarray}
\lefteqn{\Psi(R)=\int e^{-v_{0}\tau V(R)}e^{-\frac{S_{1}^{2}}{4Dt_{1}\tau}}
e^{-v_{1}\tau W(R+S_{1})} e^{-\frac{S_{2}^{2}}{4Dt_{2}\tau}}} \nonumber \\    
& & \times \,  e^{-v_{1}\tau W(R+S_{1}+S_{2})} 
e^{-\frac{(R+S_{1}+S_{2}-S_{3})^{2}}{4Dt_{1}\tau}}
e^{-v_{0}\tau V(S_{3})} \nonumber \\ 
& & \times \, \Psi_{\text m}(R+S_{1}+S_{2}+S_{3})\: dS_{1}dS_{2}dS_{3}
\label{eq:mixta}
\end{eqnarray}
 Whilst Eqs. (\ref{eq:abs}) and (\ref{eq:mixta}) are equivalent, they
lead to different estimators for the kinetic energy. 
The action of the kinetic operator on $\Psi(R)$ in Eq. (\ref{eq:mixta})
involves derivatives of $\Psi_{\text m}$, in contrast to what happens when
the prescription given in Eq. (\ref{eq:abs}) is used.

It is possible to use the propagator $G(R,S,\tau)$ in order to improve
the quality of any given trial wave function, with reduced variance
in direct proportion to its quality. In fact, Eq. (\ref{eq:mixta}) satisfies
the principle of zero variance: as the trial wave function $\Psi_{\text m}$
approaches
the exact ground-state wave function, the propagation time $\tau$
can be continuously tuned to smaller values, with the $S$ distribution
approaching a Dirac delta and $\Psi(R)$ and its derivatives approaching
the exact ones.

\begin{table*}[]
\centering
\begin{ruledtabular}
\begin{tabular}{cc|cc|cc|cc}
     &      &   \multicolumn{2}{c|}{McMillan} &
     \multicolumn{2}{c|}{Semi-classical} & \multicolumn{2}{c}{1} \\
 $\varepsilon$ (K$^{-1}$) &  $N_b$  & $E/N$ (K) & $V/N$ (K) & $E/N$ (K) &
 $V/N$ (K) & $E/N$ (K) & $V/N$ (K) 
  \\
\hline
0.01   &  1  & -6.860(58) & -20.924(30) & -3.878(76) & -17.246(44) &
-2.81(9)  & -17.009(52)    \\
0.02   &  2  & -7.175(51) & -21.106(37) & -6.234(76) & -20.285(44) &
-5.73(12)  & -20.148(42)   \\
0.04   &  4  & -7.268(44) & -21.413(35) & -7.121(67) & -21.284(43) &
-6.93(11) &  -21.186(39)  \\
0.06   &  4  & -7.303(35) & -21.538(44) & -7.306(64) & -21.547(40) & 
-7.15(11) &  -21.570(31)  \\
0.08   &  6  & -7.299(41) & -21.534(38) & -7.290(55) & -21.583(39)   \\
0.08   &  8  &            &             &            &            &
-7.25(8)  &  -21.381(20)  \\
0.10   &  8  &            &             &            &            &
-7.33(8)  &   -21.498(12)  \\
0.12   &  10 &            &             &            &            &
-7.34(9)  &  -21.524(10)  \\
\end{tabular}
\end{ruledtabular}
\caption{Total and potential energies per particle as a function of the
imaginary time $\varepsilon$ and number of beads $N_b$ using different
models for the initial wave function $\Psi_{\text m}(R)$. }
\label{table1}
\end{table*}

\begin{figure}[b]
\centerline{
        \includegraphics[width=0.82\linewidth,angle=0]{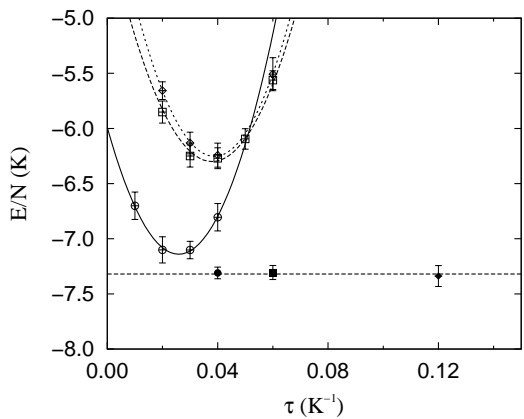}}%
        \caption{Variational energies obtained with a single time step
	$\tau$ and the proposed wave function. Open circles, squares, and
	diamonds correspond to $\Psi_{\text m}(R)$ of McMillan type,
	semiclassical model and 1, respectively. The corresponding filled
	symbols stand for the asymptotic values using the three models and
	more than one time step.  }
\label{fig1}
\end{figure}

In order to test the accuracy of the method (\ref{eq:abs},\ref{eq:mixta}),
we have applied it to liquid $^4$He, a classical benchmark in quantum
many-body physics. The calculations have been carried out at the
experimental equilibrium density $\rho=0.365\: \sigma^{-3}$ ($\sigma=2.556$
\AA) with $N=64$ atoms and using the HFD-B(HE) Aziz
potential~\cite{aziz}, which has proven to be highly accurate in the description of the
experimental equation of state~\cite{boro}. 
As a trial wave function, we use a simple Jastrow form
based on the McMillan model,
$\Psi_{\text m}(R)=\prod_{ij} \exp[- 0.5 (b/r_{ij})^{5}]$ with 
$b=1.20\:\sigma$,
and two additional variational parameters, $t_{1}$ and $\tau$ (\ref{eq:abs}). 
After a quick search
we found the optimal value $t_{1}=0.35$ which was kept fixed for
the rest of the calculations. The dependence of the variational
energy on the remaining variational parameter $\tau$ is shown in
Fig. \ref{fig1}, where the variational character of the calculation is clear.
The horizontal axis corresponds to the $\tau$ parameter in K$^{-1}$
units and the vertical axis represents the total energy per particle in K. The
data displayed as empty circles is the variational energy for the
wave function based on the high-order action (\ref{eq:abs}). 
The position of the minimum is a
compromise between a large $\tau$ value suitable for a large suppression
of the excited components present in $\Psi_{\text m}(R)$, and a small one
suitable for a proper behavior of the variational wave function (\ref{eq:abs}), 
which relies itself on a short-time expansion.
The minimum is located at $\tau=0.025$ and the variational energy
obtained is $E=-7.10$ K, only $\sim 0.2$ K higher than the exact value.

Equally impressive
is the data displayed as empty diamonds, which is a variational calculation
using the same propagator (\ref{eq:abs}) acting on $\Psi_{\text m}(R)=1$,
i.e. it is a variational calculation in which only the Hamiltonian
and the statistics are used. We see in this case that the compromise between a large
suppression of the non-ground-state components and a small $\tau$
parameter has been shifted towards a higher value $\tau=0.04$, as
could be expected. A third model for  $\Psi_{\text m}(R)$ consists in a
semi-classical approximation, $\Psi_{\text{m-sc}}(R) = \prod_{i<j} \exp[-
\varepsilon V_{\text{sr}}(r_{ij}]$, with $V_{\text{sr}}(r)= 1/r^{12}$
corresponding to the $r$-dependence of the Lennard-Jones potential around
the core.
This third model does not contain any free variational parameter since
$\varepsilon$ is the total imaginary time of the propagator. The results
obtained with $\Psi_{\text{m-sc}}(r)$ (open squares) only improve slightly
the variational energy obtained with  $\Psi_{\text m}(R)=1$.


It is also possible to apply several times the propagator 
to $\Psi_{\text m}(R)$ in order to obtain
better variational results and finally projecting out any excitation present in the model
wave functions. This is shown in Fig. \ref{fig1} with a filled circle point,
which corresponds to the propagator applied $N_{b}=4$ times
onto the Jastrow-McMillan wave function $\Psi_{\text m}(R)$ for
a total time propagation $\varepsilon=0.04$ K$^{-1}$.
The same figure shows with a filled diamond the result of acting with $N_{b}=10$ 
propagators on $\Psi_{m}(R)=1$ for  a total time
propagation $\varepsilon=0.12$ K$^{-1}$. 
Between these two points, the semiclassical model achieves
convergence very fast with $N_b=4$ and time $\varepsilon=0.06$ K$^{-1}$.


The high-order Green's function can also be used to obtain  
the total energy following the standard procedure of evaluating
the local energy of the trial wave function $\Psi_{\text m}(R)$  
at one end of the
chain, and its high accuracy translates again in the need for a very
small number of beads until convergence is achieved. 
The results obtained for the total and potential energies per particle are
reported in Table \ref{table1} as a function of the total imaginary time
$\varepsilon$. The potential energies are calculated in the center of the
chain to remove any possible bias coming from $\Psi_{\text m}(R)$; the
total energies are estimated in the extreme, except for the case $\Psi_{\text m}(R)=1$ where they
are sampled in the center.
The number
of beads for a given time $\varepsilon$ is determined simply by the requirement 
that doubling its
number (and simultaneously halving the propagation time per bead $\tau$)
does not have any effect, which turns to be equivalent to keep the
propagation time per bead below $\tau=0.15$ K$^{-1}$. One would expect
that this regime corresponds to keeping the finite time step error
just below the detectable level. This is in accordance
with Fig. \ref{fig1}, where we see that a propagation time per bead near
$0.15$ K$^{-1}$
is close to the maximum value before the time step error 
starts to become apparent by bending upwards the variational energy
curve. When both requirements: $\tau=\varepsilon/N_{b}$ to be small enough 
($\tau \leq 0.15$ K$^{-1}$ in
our case) and $N_{b}$ large enough are met, the energy becomes independent
of both $\varepsilon$ and $N_{b}$ and a good estimation of the ground-state
energy is achieved. 


The results of Table \ref{table1} show that the convergence is quickly achieved 
with only a few number
of beads: $N_b=4$ for the McMillan and semiclassical Jastrow factors, and
$N_b=8$  for $\Psi_{\text m}(R)=1$. Concerning the convergence for
the potential energy, one can see in Table \ref{table1} that the exact
(asymptotic in $\varepsilon$) value is independent 
of the trial wave function and its value is reached with only $N_b=4$. 
 It should be taken into account however that, as in any PIGS method, 
the total length of the chain corresponding
to $\left|\Psi_{\text m}(R)\right|^{2}$ is twice that of $\Psi_{\text m}(R)$,
and that one $\tau$ propagator (\ref{eq:abs}) involves three internal
\textit{shadows}.

\begin{figure}[t]
\centerline{
        \includegraphics[width=0.8\linewidth,angle=0]{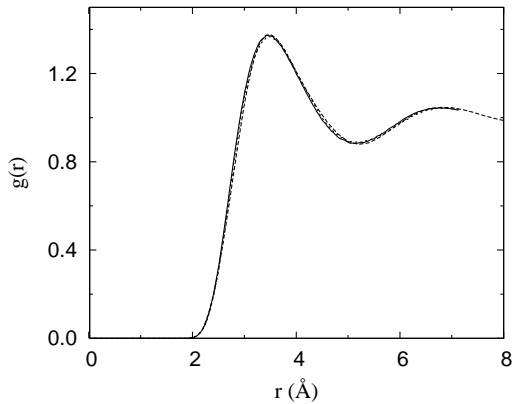}}%
        \caption{Two-body radial distribution function $g(r)$. The solid
	and dotted lines correspond to present results using for 
	$\Psi_{\text m}(R)$ a Jastrow-McMillan factor or $\Psi_{\text
	m}(R)=1$, respectively.
	The dashed line is the experimental data from Ref. \cite{svenson}}
\label{fig3}
\end{figure}

Unbised (pure) estimations of operators $\hat{O}$ other than the Hamiltonian can only
be calculated in the center of the chain,
$ \langle \hat{O} \rangle = {\cal N}^{-1} \langle \Psi_{\text m} |
G(\varepsilon/2) \hat{O}
G(\varepsilon/2) | \Psi_{\text m} \rangle$.
%
This holds, for instance, for the potential energy reported in Table \ref{table1} and the
two-body radial distribution function $g(r)$ shown in Fig. \ref{fig3}. The
present PIGS results for $g(r)$ show an excellent agreement with
experimental data~\cite{svenson} for the trial wave functions used in this work. 
Even when
$\Psi_{\text m}(R)=1$  the result is the same, the calculation requiring
only a few more beads. 

\begin{figure}[t]
\centerline{
        \includegraphics[width=0.8\linewidth,angle=0]{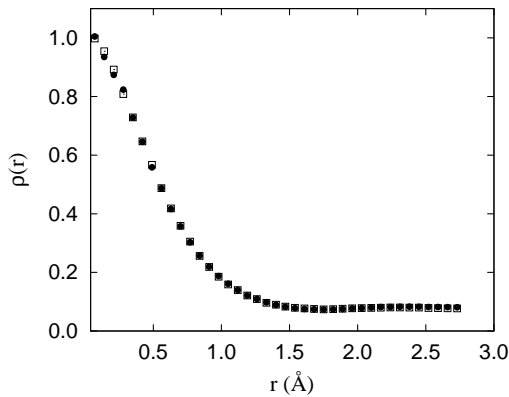}}%
        \caption{One-body density matrix $\rho(r)$. Filled circles and open
	squares stand for PIGS results using for 
	$\Psi_{\text m}(R)$ a Jastrow-McMillan factor ($N_b=5$) or
	$\Psi_{\text m}(R)=1$
	($N_b=10$), respectively. The error bars are smaller than the size
	of the symbols.}
\label{fig4}
\end{figure}

Another relevant function that can be computed in an unbiased way is the
one-body density matrix $\rho(r)$, whose asymptotic limit is the condensate
fraction. The calculation of $\rho(r)$ has been carried out by
incorporating \textit{worm} movements in the sampling~\cite{worm}, 
a technique that has been developed 
for path integral Monte Carlo (finite temperature) and that we have
extended to PIGS. One of the main advantages of this method is 
that $\rho(r)$ comes properly normalized, and thus 
eliminates any uncertainty introduced by the
a posteriori normalization factor. In Fig. \ref{fig4}, results for
$\rho(r)$ obtained using different trial wave functions $\Psi_{\text m}$ are shown. 
As one can see,
the results are statistically indistinguishable and predict a condensate
fraction $n_0=0.080(2)$, in nice agreement with a recent PIMC estimation at
$T=1$ K ($n_0=0.081(2)$)~\cite{worm}.

Summarizing, the high-order
short time expansion of the Green's function presented in this work
enables the possibility of performing variational calculations
of very high-quality on systems for which no model of wave function is known.
This is illustrated in the case of liquid $^{4}$He at equilibrium
density, where the propagator
provides for $N_{b}=1$  and $\Psi_{\text m}(R)=1$  a variational energy
$E/N=-6.20$ K,
and converges to the exact value already with $N_{b}=8$.
When used to improve a Jastrow-McMillan wave function, it produces
a shadow-like variational wave function for $N_{b}=1$ with a variational
ground-state energy $E/N=-7.10$ K, only $\sim 0.2$ K higher than the exact
value. Repeated application of the
propagator leads to variational results which are
asymptotically exact for values as low as $N_{b}=4$. 
The prospects for future work are promising, since this opens the
road to being able to obtain results for systems whose ground-state
wave function is poorly known or even unknown.

We wish to thank stimulating discussions with Siu Chin.
Partial financial support from DGI (Spain) Grant No.
 FIS2008-04403 and Generalitat de Catalunya Grant No. 2005SGR-00779 is also
 akcnowledged.

\end{document}